\documentclass[conference]{IEEEtran}
\IEEEoverridecommandlockouts

\usepackage[utf8]{inputenc}
\usepackage[T1]{fontenc}
\usepackage{pgf}
\usepackage{cite}
\usepackage{amsmath,amssymb,amsfonts}
\usepackage{algorithmic}
\usepackage{graphicx}
\usepackage{textcomp}
\usepackage{xcolor}
\usepackage{notation}
\usepackage{multirow}
\usepackage{acronym}
\usepackage{psfrag}
\usepackage{soul}
\usepackage{import}
\usepackage[inline]{enumitem}
\usepackage[level=3]{LatexInclusion/messages}
\usepackage{auto-pst-pdf}
\usepackage{tikz}

\acrodef{bp}[BP]{belief propagation}
\acrodef{slam}[SLAM]{simultaneous localization and mapping}
\acrodef{2d}[2-D]{two dimensional}
\acrodef{po}[PO]{potential object}
\acrodef{mmse}[MMSE]{minimum mean square error}
\acrodef{rmse}[RMSE]{root mean square error}
\acrodef{pdf}[PDF]{probability density function}
\acrodef{pmf}[PMF]{probability mass function}
\acrodef{gospa}[GOSPA]{generalized optimal sub-pattern assignment}
\acrodef{roi}[ROI]{region of interest}
\acrodef{fft}[FFT]{fast Fourier transform}
\acrodef{sbl}[SBL]{sparse Bayesian learning}
\acrodef{mp}[MP]{matching pursuit}
\acrodef{vla}[VLA]{vertical linear array}
\acrodef{iid}[i.i.d.]{independent and identically distributed}
\acrodef{tbd}[TBD]{track-before-detect}
\acrodef{mot}[MOT]{multiobject tracking}
\acrodef{snr}[SNR]{signal-to-noise ratio}

\newcommand{\ist}{\hspace*{.3mm}}
\newcommand{\rmv}{\hspace*{-.3mm}}

\newcommand{\nn}{\nonumber}
\newcommand{\T}{\mathrm{T}}
\newcommand{\CH}{\mathrm{H}}

\def\BibTeX{{\rm B\kern-.05em{\sc i\kern-.025em b}\kern-.08em
    T\kern-.1667em\lower.7ex\hbox{E}\kern-.125emX}}
\begin{document}

\title{An Approach of Directly Tracking \\ Multiple Objects}

\author{\IEEEauthorblockN{Mingchao Liang and Florian Meyer \\ [1.2mm]}
\IEEEauthorblockA{Department of Electrical and Computer Engineering, University of California San Diego, USA \\
\texttt{\{m3liang, flmeyer\}@ucsd.edu}}
}

\maketitle


\begin{abstract}
In conventional approaches for \ac{mot}, raw sensor data undergoes several preprocessing stages to reduce data rate and computational complexity. This typically includes coherent processing that aims at maximizing the \ac{snr}, followed by a detector that extracts ``point'' measurements, e.g., the range and bearing of objects, which serve as inputs for sequential Bayesian \ac{mot}. While using point measurements significantly simplifies the statistical model, the reduced data rate can lead to a loss of critical, object-related information and, thus, potentially to reduced tracking performance. In this paper, we propose a direct tracking approach that avoids a detector and most preprocessing stages. For direct tracking, we introduce a measurement model for the data-generating process of the sensor data, along with state-transition and birth models for the dynamics and the appearance and disappearance of objects. Based on the new statistical model, we develop a factor graph and particle-based \ac{bp} method for efficient sequential Bayesian estimation. Contrary to the \ac{tbd} paradigm which also avoids a detector, direct tracking integrates coherent processing within the Bayesian \ac{mot} framework. Numerical experiments based on a passive acoustic dataset demonstrate that the proposed direct approach outperforms state-of-the-art conventional methods that rely on multiple preprocessing stages\vspace{3mm}.
\end{abstract}


\section{Introduction}

Accurately detecting and tracking objects in the environment is an essential yet challenging task, with many applications ranging from applied ocean science to autonomy in robotics \cite{BarWilTia:B11,Mah:B07,ChaMor:B11,MeyKroWilLauHlaBraWin:J18,LiaMey:J23}. Conventional methods for \ac{mot} rely on a preprocessing stage first applied to the raw data. This preprocessing stage typically consists of a bank of matched filters for coherent processing and maximization of the \ac{snr}. This coherent processing in time and space leads to a grid of data cells. Based on an object detector \cite{Nit:J72,MalZha:J93,NanGemGerHodMec:J19,YinZhoKra:21}, data cells are then converted to so-called ``point measurements'' used as the input for Bayesian \ac{mot}. The detector used in such conventional MOT approaches reduces data flow and computational complexity. However, object-related information may be lost during this preprocessing stage, leading to reduced tracking performance. In addition, the performance of detectors can be constrained by low-resolution sensors, creating additional bottlenecks for tracking accuracy.

A line of research that aims at reducing the information loss during preprocessing, led to the development of \ac{tbd} methods for the tracking of a single \cite{TonBar:J98,KimRisGuaRos:21} or multiple objects, \cite{VoVoPhaSut:10,LepRabGla:J16,KroWilMey:21,LiaKroMey:J23,KroWilMey:C24,DavGar:J24}. \Ac{tbd} is typically directly applied to the data cells, avoiding any object detector. In \ac{tbd}, potentially suboptimal decisions on the size and number of data cells have to be made, and the influence of target states on data cells is determined by so-called point spread functions \cite{TonBar:J98,KimRisGuaRos:21}, which are often difficult to model probabilistically. Oversimplified point spread functions, e.g., Gaussian kernels, can introduce model mismatch and degrade the tracking performance.


In this paper, we propose a Bayesian \ac{mot} method that can potentially be applied to raw sensor data. Our approach can avoid the formation of data cells by performing coherent processing across sensor elements within the \ac{mot} method. In the considered sequential Bayesian setting, our direct tracking method aims to compute posterior PDFs of object states, which, at each discrete time step, consist of a kinematic state, a binary existence variable, and transmit power. We introduce a measurement model that captures the statistical relationship between sensor data and the object states. Here, the signal amplitude is assumed random and zero-mean, Gaussian distributed with a variance governed by the object's existence variable and transmit power. The resulting hierarchical Bernoulli-Gaussian model \cite{HanFleRao:J18} has sparsity-promoting features that facilitate the separation of signal contributions of closely spaced objects. We also introduce statistical models for both the dynamics and the birth of objects, which mostly follow existing models in the \ac{mot} literature \cite{Wil:J15,MeyKroWilLauHlaBraWin:J18}. 

Based on the joint statistical model, we develop a factor graph that enables the development of a particle-based \ac{bp} method \cite{KscFreLoe:01, Loe:04} for efficient inference of marginal posterior PDFs. While our approach is very general and can be applied to a wide range of active and passive tracking problems, we here focus on the passive tracking of sources that emit tonal signals. The main contributions of this paper are as follows\vspace{.5mm}:
\begin{itemize}
    \item We introduce a multisnapshot and multifrequency measurement model that characterizes the data generation process of the sensor data in as passive tracking problem\vspace{2mm}.
    \item We represent the statistical models for direct tracking by a factor graph and develop a computationally efficient \ac{bp} method for inference\vspace{2mm}. 
    \item We evaluate our proposed method using a challenging passive acoustic dataset and show that it has a superior performance compared to conventional methods.
    \vspace{1mm}
\end{itemize}

\pagebreak
Unlike conventional \ac{mot} methods with multiple preprocessing stages, our proposed method has the potential to leverage the full information from the sensor data, resulting in an improved \ac{mot} performance.





\section{System Model} \label{sec:sig}

The signal modeling of the considered object tracking problem is introduced as follows. At each time step $k$, there are $L_k$ objects, each transmitting $I$ tones at frequencies $f_i, i \in \{1, \dots, I\}$ simultaneously. The kinematic state $\V{x}_{k, l}$ of object $l \in \{1\dots,L_k\}$ includes the object's position and velocity. Multiple measurement vectors, also referred to as snapshots, are received for each tone $i \in \{1, \dots, I\}$ by the sensor. The $j$-th measurement vector $\V{z}_{k, j}^{(i)} \in \mathbb{C}^M, i \in \{1, \dots, I\}, j \in \{1, \dots, J\}$ can be modeled as  
\begin{equation}
 \label{eq:signal_model} 
    \V{z}_{k, j}^{(i)} = \sum_{l = 1}^{L_k} \varrho_{k, l, j}^{(i)} \V{a}^{(i)}(\V{x}_{k, l}) + \V{\epsilon}_{k, j}^{(i)}
\end{equation}
where $\V{a}^{(i)}(\V{x}_{k, l})$ is a known function that describes the contribution of a target with state $\V{x}_{k, l}$ to the measurement vector $\V{z}_{k, j}^{(i)}$. Furthermore, $\varrho_{k, l, j}^{(i)} \in \mathbb{C}$ is the complex amplitude of the $i$-th tone, related to object $l$ at snapshot $j$, and $\V{\epsilon}_{k, j}^{(i)} \in \mathbb{C}^M$ is the noise vector. The parameters $L_k, \varrho_{k, l, j}^{(i)},$ and $\V{x}_{k, l}$ are assumed unknown.

\subsection{Measurement Model} \label{subsec:meas_model}

To represent the unknown and time-varying number of objects, $L_k$, more suitably for estimation, we introduce $N_k$ \acp{po} \cite{MeyKroWilLauHlaBraWin:J18} with kinematic states $\V{x}_{k, n}, n \in \{1, \dots, N_k\}$. Each \ac{po} $n \in \{1, \dots, N_k\}$ has an associated binary random variable $r_{k, n} \in \{0, 1\}$ which indicates its existence. In particular, $r_{k, n} = 1$ indicates that \ac{po} $n$ exists; while $r_{k, n} = 0$ indicates that it does not exist. By introducing \acp{po}, the signal model in \eqref{eq:signal_model} can be expressed\vspace{-1mm} as 


\begin{equation}
    \V{z}_{k, j}^{(i)} = \sum_{n = 1}^{N_k} r_{k, n} \rho_{k, n, j}^{(i)} \V{a}_{k, n}^{(i)} + \V{\epsilon}_{k, j}^{(i)} \label{eq:signal_model_po}
\end{equation}
where $\V{a}_{k, n}^{(i)} \triangleq \V{a}^{(i)}(\V{x}_{k, n})$ is the contribution vector from \ac{po} $n \in \{1, \dots, N_k\}$ to the measurement at tone $i \in \{1, \dots, I\}$. The complex amplitude $\rho_{k, n, j}^{(i)}$ is assumed zero-mean complex Gaussian with a common variance $\gamma_{k, n}^{(i)}$ across snapshots, i.e., $\rho_{k, n, j}^{(i)} \sim \mathcal{CN}(\rho_{k, n, j}^{(i)}; 0, \gamma_{k, n}^{(i)}), \forall j \in \{1, \dots, J\}$. We define $\V{\phi}_{k, n} = [\V{\gamma}_{k, n}^\T \hspace{1mm} r_{k, n}]^\T$ as the power state of \ac{po} $n$, where $\V{\gamma}_{k, n} = [\gamma_{k, n}^{(1)} \cdots \gamma_{k, n}^{(I)}]^\T$ represents transmit power. The amplitudes are assumed mutually independent and are independent of all $\V{x}_{k, n}$, $r_{k, n}$, and $\V{\epsilon}_{k, j}^{(i)}$. The noise vector $\V{\epsilon}_{k, j}^{(i)}$ is also assumed to be white, zero-mean complex Gaussian with a common power for all snapshots, i.e., $\V{\epsilon}_{k, j}^{(i)} \sim \mathcal{CN}(\V{\epsilon}_{k, j}^{(i)}; \V{0}, \eta_{k}^{(i)} \M{I}_M), \forall j \in \{1, \dots, J\}$. Conditioned on noise power, the measurement noise is furthermore mutually independent across tones and snapshots and independent of all $\V{x}_{k, n}$, $r_{k, n}$, and $\rho_{k, n, j}^{(i)}$. Based on the above assumptions, the likelihood function of individual measurement vectors, $\V{z}_{k, j}^{(i)}$, is zero-mean complex Gaussian and given by 
\begin{align}
    f(\V{z}_{k, j}^{(i)} | \V{x}_k, \V{\phi}_{k}, \eta_{k}^{(i)}) = \mathcal{CN}(\V{z}_{k, j}^{(i)}; \V{0}, \M{C}_{k}^{(i)}) \nonumber
\end{align}
with covariance $\M{C}_{k}^{(i)} = \sum_{n = 1}^{N_k} r_{k, n} \gamma_{k, n}^{(i)}  \V{a}_{k, n}^{(i)} \V{a}_{k, n}^{(i) \CH} + \eta_{k}^{(i)} \M{I}_M$. Here, we introduced $\V{x}_k = [\V{x}_{k , 1}^\T \cdots \V{x}_{k, N_k}^\T ]^\T$ and $\V{\phi}_{k} = [\V{\phi}_{k, 1}^\T \cdots \V{\phi}_{k, N_k}^{\T}]^\T$\hspace{-.3mm}. The joint likelihood function can also be obtained as the product of individual likelihood functions, i.e., 
\begin{equation}
    f(\V{z}_{k} | \V{x}_k, \V{\phi}_{k}, \V{\eta}_{k}) = \prod_{i = 1}^I \prod_{j = 1}^J f(\V{z}_{k, j}^{(i)} | \V{x}_k, \V{\phi}_{k}, \eta_{k}^{(i)}) \nn
\end{equation}
with $\V{z}_{k} = [\V{z}_{k}^{(1) \T}  \cdots \V{z}_{k}^{(I) \T} ]^\T$\hspace{-.3mm}, $\V{z}_{k}^{(i)} = [\V{z}_{k, 1}^{(i) \T} \cdots \V{z}_{k, J}^{(i) \T} ]^\T$\hspace{-.3mm}, and $\V{\eta}_{k} = [\eta_{k}^{(1)} \cdots \eta_{k}^{(I)}]^\T$\hspace{-.3mm}. For future reference, we introduce the joint measurement vector $\V{z}_{1 : k} = [\V{z}_{1}^\T \cdots \V{z}_{k}^\T]^\T$\hspace{-.3mm}.



\subsection{State-Transition and Birth PDFs} \label{subssec:state_tran_model}

It is assumed that the state of each \ac{po} evolves independently. The dynamics of kinematic state $\V{x}_{k, n}, n \in \{1, \dots, N_{k - 1}\}$, are described by the conditional \ac{pdf} $f(\V{x}_{k, n} | \V{x}_{k - 1, n})$ which can, e.g., follow a constant velocity model \cite[Ch. 4]{ShaKirLi:B02}. The dynamics of the power state $\V{\phi}_{k, n}, n \in \{1, \dots, N_{k - 1}\}$, are described by the conditional \ac{pdf} $f(\V{\phi}_{k, n} | \V{\phi}_{k - 1, n}) = p(r_{k, n} | r_{k - 1, n}) f(\V{\gamma}_{k, n} | \V{\gamma}_{k - 1, n})$, where we assume that existence and power evolve independently. If a \ac{po} does not exist at the previous time step $k - 1$, then it also does not exist at time $k$, i.e., $p(r_{k, n} = 1 | r_{k, n - 1} = 0) = 0$. If it exists at time $k - 1$, then it continues to exist at time $k$ with survival probability $p_{\mathrm{s}}$, i.e., $p(r_{k, n} = 1 | r_{k, n - 1} = 1) \rmv=\rmv p_{\mathrm{s}}$. For each tone $i \in \{1, \dots, I\}$, the measurement noise variances are assumed to be independent, i.e., $f(\V{\eta}_{k} | \V{\eta}_{k - 1}) = \prod_{i = 1}^{I} f(\eta_{k}^{(i)} | \eta_{k - 1}^{(i)})$ and  the individual state-transition \acp{pdf} $f(\eta_{k}^{(i)} | \eta_{k - 1}^{(i)})$ are assumed to follow a Gamma distribution.

To model newly appearing objects, at each time step $k$, we introduce $Q$ new \acp{po} in the \ac{roi}. The ROI is equal to the support $\Set{X}$ of kinematic states $\V{x}$. The birth of newly appearing objects is modeled by a Poisson point process with mean $\mu_{\text{B}}$ and spatial \ac{pdf} $f_{\text{B}}(\V{x}, \V{\gamma}) = f_{\text{B}}(\V{x}) f_{\text{B}}(\V{\gamma})$, where we assume the prior of kinematic states and powers of new \acp{po} are independent, and $f_{\text{B}}(\V{x}) = 0$ for $\V{x} \not \in \Set{X}$. The $Q$ new \acp{po} occupy $Q$ non-overlapping regions that partition $\Set{X}$. The non-overlapping regions are denoted by $\Set{X}_q, q \in \{1, \dots, Q\}$ with $\uplus_{q = 1}^Q \Set{X}_q = \Set{X}$ and corresponding new \acp{po} are indexed as $n = N_{k - 1} + q$, with $q \in \{1, \dots, Q\}$. As a result of this partitioning, the birth process within each region $\Set{X}_q$, is also a Poisson point process with mean $\mu_{\text{B}, n} = \mu_{\text{B}} \int_{\Set{X}_q} f_{\text{B}}(\V{x}) \hspace{1mm} \mathrm{d} \V{x}$ and spatial \ac{pdf} $f_{\text{B}, n}(\V{x})$.  If $\V{x} \in \Set{X}_q$, this spatial \ac{pdf} is equal to  $f_{\text{B}}(\V{x})$, up to a normalization constant, and zero otherwise. By making the region covered by each $\Set{X}_q$ sufficiently small, we can assume that there is at most one object newly appearing in $\Set{X}_q$. The birth probability of each new \ac{po} can thus be set to $p_{\text{B}, n} = \mu_{\text{B}, n} / (\mu_{\text{B}, n} + 1)$ and the \ac{pdf} of a new \ac{po} $n \in \{N_{k - 1} + 1, \dots, N_k\}$ is given by $f(\V{x}_{k, n}, \V{\phi}_{k, n}) = f(\V{x}_{k, n}) f(\V{\phi}_{k, n})$ with $f(\V{x}_{k, n})$ and $f(\V{\phi}_{k, n})$ defined\vspace{-1mm} as
\begin{align}
    f(\V{x}_{k, n}) &= f_{\text{B}, n}(\V{x}_{k, n}) \hspace{4mm} \nn \\
    f(\V{\phi}_{k, n}) &= p_{\text{B}}(r_{k, n}) f_{\text{B}}(\V{\gamma}_{k, n}) \nn
\end{align}
and $p_{\text{B}}(r_{k, n} = 1) = p_{\text{B}, n}$.

To facilitate an efficient computation of the marginal posterior \acp{pdf}, we introduce further conditional independence assumptions as follows. For all the state-transition models, we also assume that, conditioned on $\V{x}_{k-1, n}, n \in \{1, \dots, N_{k - 1}\}$, the kinematic state $\V{x}_{k, n}$ is independent of all current and previous power states, all current and previous measurement noise variances, and the current and previous kinematic states of all other \acp{po}. Similarly, conditioned on $\V{\phi}_{k-1, n}, n \in \{1, \dots, N_{k - 1}\}$, the power state $\V{\phi}_{k, n}$ is independent of all current and previous kinematic states, all current and previous measurement noise variances, and the current and previous\vspace{.4mm} power states of all other \acp{po}. The noise variance $\eta_{k}^{(i)}$ is also assumed to be independent of current and previous states of all \acp{po}, conditioned on $\eta_{k - 1}^{(i)}$. For the birth model, we also assume that the kinematic and power states $\V{x}_{k, n}, \V{\phi}_{k, n}, n \in \{N_{k - 1} + 1, \dots, N_k\}$ of new \acp{po} are independent of all current and previous kinematic and power states of all other \acp{po}.

Finally, at time $k = 0$, the prior distributions $f(\V{x}_{0, n})$, $f(\V{\phi}_{0, n})$, $n \in \{1, \dots, N_0\}$, and $f(\eta_{0}^{(i)}), i \in \{1, \dots, I\}$ are assumed known and the initial random parameters $\V{x}_{0, n}, \V{\phi}_{0, n}, n \in \{1, \dots, N_0\}$ and $\eta_{0}^{(i)}, i \in \{1, \dots, I\}$ are all independent of each other.

\begin{figure}[!tbp]
    \centering
    \psfrag{da1}[c][c][0.75]{\raisebox{-2mm}{\hspace{.8mm}$\V{x}_{1}$}}
    \psfrag{daI}[c][c][0.75]{\raisebox{-2.5mm}{\hspace{.3mm}$\V{x}_{\underline{N}}$}}
    \psfrag{db1}[c][c][0.75]{\raisebox{-2mm}{\hspace{.2mm}$\V{x}_{\scriptscriptstyle \underline{N} + 1}$}}
    \psfrag{dbJ}[c][c][0.75]{\raisebox{1mm}{$\V{x}_{N}$}}
    \psfrag{q1}[c][c][0.75]{\raisebox{-1mm}{$f_{1}^{\V{x}}$}}
    \psfrag{qI}[c][c][0.75]{\raisebox{-3mm}{$f_{\underline{N}}^{\V{x}}$}}
    \psfrag{v1}[c][c][0.75]{\raisebox{-3.3mm}{\hspace{.15mm}$f_{\scriptscriptstyle \underline{N} + 1}^{\V{x}}$}}
    \psfrag{vJ}[c][c][0.75]{\raisebox{-1mm}{\hspace{.3mm}$f_{N}^{\V{x}}$}}
    \psfrag{g1}[c][c][0.75]{\raisebox{-1mm}{\hspace{.3mm}$f_{1}^{\V{z}}$}}
    \psfrag{gJ}[c][c][0.75]{\raisebox{-1.7mm}{\hspace{.5mm}$f_{J}^{\V{z}}$}}
    \psfrag{ma1}[c][c][0.75]{\color{blue}{$\alpha_{1}$}}
    \psfrag{maJ}[l][l][0.75]{\color{blue}{$\alpha_{N}$}}
    \psfrag{mb11}[l][l][0.75]{\color{blue}{$\beta_{1, 1}$}}
    \psfrag{mk11}[r][r][0.75]{\color{blue}{\raisebox{1.5mm}{$\kappa_{1, 1}$}}}
    \psfrag{fn}[c][c][0.75]{$f^{\eta}$}
    \psfrag{dnI}[c][c][0.75]{$\eta$}
    \psfrag{me}[c][c][0.75]{\color{blue}{$\xi$}}
    \psfrag{mxiJ}[c][c][0.75]{\raisebox{-3mm}{\color{blue}{\hspace{6mm}$\iota_J$}}}
    \psfrag{mnuJ}[l][l][0.75]{\raisebox{-3mm}{\color{blue}{\hspace{0mm}$\nu_J$}}}
    \psfrag{dc1}[c][c][0.75]{\raisebox{-2mm}{\hspace{.8mm}$\V{\phi}_{1}$}}
    \psfrag{dcI}[c][c][0.75]{\raisebox{-2.5mm}{\hspace{.3mm}$\V{\phi}_{\underline{N}}$}}
    \psfrag{dd1}[c][c][0.75]{\raisebox{-2mm}{\hspace{.2mm}$\V{\phi}_{\scriptscriptstyle \underline{N} + 1}$}}
    \psfrag{ddJ}[c][c][0.75]{\raisebox{1mm}{$\V{\phi}_{N}$}}
    \psfrag{qp1}[c][c][0.75]{\raisebox{-1mm}{$f_{1}^{\V{\phi}}$}}
    \psfrag{qpI}[c][c][0.75]{\raisebox{-3mm}{$f_{\underline{N}}^{\V{\phi}}$}}
    \psfrag{vp1}[c][c][0.75]{\raisebox{-3.3mm}{\hspace{.15mm}$f_{\scriptscriptstyle \underline{N} + 1}^{\V{\phi}}$}}
    \psfrag{vpJ}[c][c][0.75]{\raisebox{-1mm}{\hspace{.3mm}$f_{N}^{\V{\phi}}$}}
    \psfrag{mp1}[c][c][0.75]{\color{blue}{$\psi_{1}$}}
    \psfrag{ms11}[r][r][0.75]{\color{blue}{$\lambda_{1, 1}$}}
    \psfrag{mz11}[l][l][0.75]{\color{blue}{\raisebox{1.0mm}{$\zeta_{1, 1}$}}}
    \includegraphics[scale=0.8]{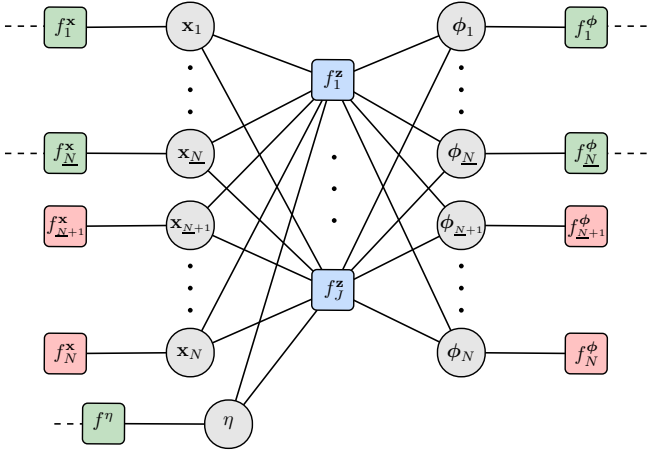}
    \caption{Factor graph representing the joint posterior distribution in \eqref{eq:factorization} for a single time step $k$ and a single tone $I\rmv=\rmv1$. The time index $k$ and the tone index $i$ are omitted. The following shorthand notations are used: $\underline{N} = N_{k - 1}$, $N = N_k$, $\V{x}_n = \V{x}_{k, n}$, $\V{\phi}_n = \V{\phi}_{k, n}$, $\eta = \eta_k^{(i)}$, $f_{j}^{\V{z}} = f(\V{z}_{k, j}^{(i)} | \V{x}_k, \V{\phi}_{k}, \eta_{k}^{(i)})$, $f_{n}^{\V{x}} = f(\V{x}_{k, n} | \V{x}_{k - 1, n})$, $f_{n}^{\V{\phi}} = f(\V{\phi}_{k, n} | \V{\phi}_{k - 1, n})$ for $n \in \{1, \dots, N_{k - 1}\}$,  $f_{n}^{\V{x}} = f(\V{x}_{k, n})$, $f_{n}^{\V{\phi}} = f(\V{\phi}_{k, n})$ for $n \in \{N_{k - 1} + 1, \dots, N_k\}$.}
    \label{fig:factor_graph}
\end{figure}

\section{Proposed Method} \label{sec:bp}

In this section, we introduce the factor graph related to the statistical model in Sec.~\ref{sec:sig} and briefly discuss our method for solving the considered \ac{mot} problem. Since the existence of \acp{po} is modeled by binary random variables, $r_{k, n}$, estimating the number of objects can be performed by (i) computing posterior existence probabilities $p(r_{k , n} | \V{z}_{1 : k})$ and  (ii) comparing these probabilities with a threshold $T_{\mathrm{dec}}$ to declare the existence of \acp{po}. Since the number of \acp{po} grows with time, to reduce the complexity, we prune a \ac{po}, i.e., remove its kinematic and power state from the state space, if its existence probability is lower than a threshold $T_{\mathrm{pru}}$. To estimate the kinematic states of \acp{po} that have been declared to exist, we aim to compute \ac{mmse} estimates, which are given by
\begin{equation}
    \hat{\V{x}}_{k, n} = \int \V{x}_{k, n} f(\V{x}_{k , n} | \V{z}_{1 : k}) \mathrm{d} \V{x}_{k, n}. \nn
\end{equation}

\Ac{mmse} estimation requires the knowledge of the marginal posterior \ac{pdf} $f(\V{x}_{k , n} | \V{z}_{1 : k})$. Both the posterior existence probabilities $p(r_{k , n} | \V{z}_{1 : k})$ and  the marginal posterior \acp{pdf} $f(\V{x}_{k , n} \ist | \ist \V{z}_{1 : k})$ are marginals of the the joint posterior \ac{pdf} $f(\V{x}_{0 : k}, \V{\phi}_{0 : k}, \V{\eta}_{0 : k} | \V{z}_{1 : k})$. Direct computation of these marginals, however, is computationally infeasible.

Provided the conditional independence assumptions as well as the statistical models in Section~\ref{sec:sig}, the joint posterior \ac{pdf}, $f(\V{x}_{k , n} \ist | \ist \V{z}_{1 : k})$, can be factorized as
\begin{align}
    &f(\V{x}_{0 : k}, \V{\phi}_{0 : k}, \V{\eta}_{0 : k} | \V{z}_{1 : k})  \nn \\
    &\propto \prod_{n = 1}^{N_0} f(\V{x}_{0, n}) f(\V{\phi}_{0, n}) \Big( \prod_{i = 1}^I f(\eta_0^{(i)}) \Big) \prod_{k' = 1}^{k} \Big( \prod_{i = 1}^I f(\eta_{k'}^{(i)} | \eta_{k' - 1}^{(i)}) \Big)\nn \\
    &\times \prod_{n = 1}^{N_{k' - 1}} f(\V{x}_{k', n} | \V{x}_{k' - 1, n}) f(\V{\phi}_{k', n} | \V{\phi}_{k' - 1, n}) \prod_{n = N_{k' - 1} + 1}^{N_{k'}} f(\V{x}_{k', n}) \nn \\
    &\times  f(\V{\phi}_{k', n}) \Big( \prod_{i = 1}^I \prod_{j = 1}^J f(\V{z}_{k', j}^{(i)} | \V{x}_{k'}, \V{\phi}_{k'}, \eta_{k'}^{(i)}) \Big) \label{eq:factorization}
\end{align}
where we have introduced $\V{x}_{0 : k} = [\V{x}_{0}^\T \cdots \V{x}_{k}^\T]^\T$, $\V{\phi}_{0 : k} = [\V{\phi}_{0}^\T \cdots \V{\phi}_{k}^\T]^\T$, and $\V{\eta}_{0 : k} = [\V{\eta}_{0}^\T \cdots \V{\eta}_{k}^\T]^\T$\hspace{-.5mm}. A single time step of the factor graph \cite{KscFreLoe:01, Loe:04} corresponding to the factorization in \eqref{eq:factorization}, is shown in Fig.~\ref{fig:factor_graph}. 

The factorization in \eqref{eq:factorization} enables the development of a \ac{bp} method \cite{KscFreLoe:01} for efficient approximate marginalization. \Ac{bp}, also known as the sum-product algorithm, is an algorithm that aims at calculating marginal posterior \acp{pdf} efficiently by utilizing the structure of the factor graph. It performs local operations on the edges of the factor graph and computes so-called ``messages'' that are real-valued functions of the unknown parameters. Within our approach, \ac{bp} messages that are computationally intractable are approximated using particles \cite{IhlMca:09} or Gaussian \acp{pdf} similarly as in \cite{LiaKroMey:J23,DavGar:J24,LiaLeiMey:J24arixv}. Following the standard sum-product rule \cite{KscFreLoe:01}, we can obtain approximate existence probability $\tilde{p}(r_{k, n})$ and approximate marginal posterior $\tilde{f}(\V{x}_{k, n})$, which can then be used for object declaration and state estimation. Details on the BP method will be provided in the journal version of this paper.



\section{Results} 
\vspace{-.5mm}

\begin{figure}[!tbp]
    \centering
    \hspace{-15mm} \resizebox{0.75\linewidth}{!}{\input{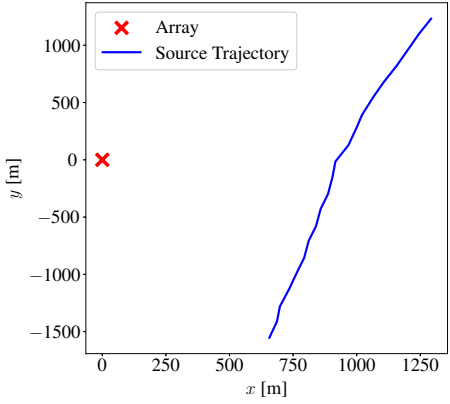}} 
    \vspace{-4mm}
    \caption{Source trajectory and the position of the acoustic array.}
    \label{fig:gt_trajectory}
    \vspace{-3.5mm}
\end{figure}

To evaluate the performance of the proposed method, we make use of a passive acoustic dataset \cite{BooAbaSchHod:J00}. In particular, we consider a 20-minute segment in the dataset, where a source moves along the trajectory shown in Fig.~\ref{fig:gt_trajectory} at around 58 m depth. The source transmits $I = 7$ tones with frequencies $f_i \in \{49, 79, 112, 148, 201, 283, 388\}$~Hz simultaneously, which are recorded by a \ac{vla} with $M = 21$ elements at a known position shown in Fig.~\ref{fig:gt_trajectory}. To illustrate the capability of the proposed method to track multiple objects,  following \cite{GemNanGerHod:J17}, we construct a scenario with two sources by adding the \ac{vla} data at a certain time step to the original \ac{vla} data. This results in a scenario with a static and a moving source transmitting at the same frequencies. The power of the moving source is scaled such that it is 3dB weaker than the power of the static one. The time between discrete time steps is 4.096 seconds, leading to a total of 292 time steps. In addition, there are $J = 3$ snapshots at each time step. Our method aims at estimating the number of sources as well as their ranges and depths from the \ac{vla} using the measurements received by the \ac{vla}.




\begin{figure}[!tbp]
    \centering
    \hspace{-5mm} \resizebox{0.9\linewidth}{!}{\input{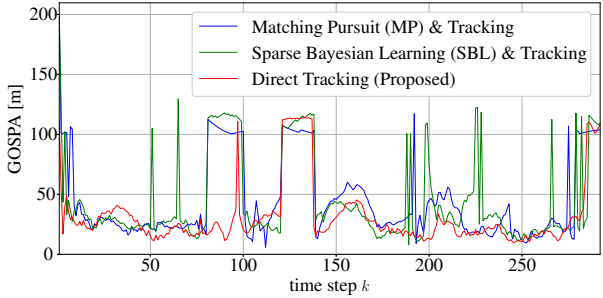}} 
    \vspace{-2mm}
    \caption{GOSPA error of the different tracking methods.}
    \label{fig:gospa}
    \vspace{-5mm}
\end{figure}

\begin{figure}[!tbp]
    \centering
    \resizebox{0.9\linewidth}{!}{\import{Figs/rawsignal/}{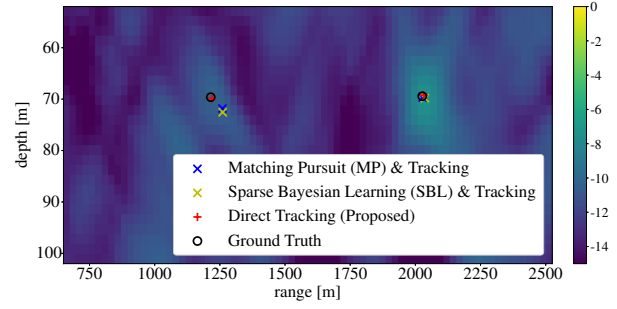}} 
    \vspace{-2mm}
    \caption{Ground truth and state estimates obtained by the different tracking methods at time step $k = 200$. The ambiguity surface provided by conventional beamforming is shown in the background.}
    \label{fig:qualitative}
    \vspace{-5mm}
\end{figure}

The kinematic state $\V{x}_{k, n} \in \mathbb{R}^4$ contains the range, depth, range rate, and depth rate. We consider a maximum range of 5000~m and a maximum depth of 200~m, leading to the \ac{roi} $\Set{X} = [0, 5000] \times [0, 200] \times \mathbb{R}^2$. For the dynamics of the kinematic states, we use the constant rate model $\V{x}_{k, n} = \M{F}\V{x}_{k - 1, n} + \M{W}\V{q}_k$ \cite[Ch. 4]{ShaKirLi:B02}, where $\V{q}_k \in \mathbb{R}^2$ is the zero-mean Gaussian driving noise  with covariance matrix $\text{diag}(10^{-2}, 10^{-5})$. The state-transition of the power for each tone is assumed to be independent, i.e., $f(\V{\gamma}_{k, n} | \V{\gamma}_{k - 1, n} ) = \prod_{i = 1}^7 f(\gamma_{k, n}^{(i)} | \gamma_{k - 1, n}^{(i)} )$, and the individual state-transition \acp{pdf} follow a Gamma distribution $f(\gamma_{k, n}^{(i)} | \gamma_{k - 1, n}^{(i)}) = \mathcal{G}(\gamma_{k, n}^{(i)}; \gamma_{k - 1, n}^{(i)} / c_{\gamma}, c_{\gamma})$ with $c_{\gamma} = 10^4$. The mean of newborn \acp{po} is set to $\mu_{\text{B}} = 10^{-4}$ and the spatial \ac{pdf} is uniform on $[0, 5000] \text{m} \times [0, 200] \text{m}$ for range and depth, and also uniform on $[-4, 4] \text{m/s} \times [-1, 1] \text{m/s}$ for range rate and depth rate. The regions $\Set{X}_q$ are created by a range and depth discretization with a resolution of 25~m and 2~m that results in a 200-by-100 grid on the \ac{roi}. The transmit powers of new \acp{po} are uniformly distributed between 0 to 1 and independent across the seven tones. Instead of initializing new \acp{po} for all grid points, we pick the top ten grid points using \ac{mp} \cite{MalZha:J93} and only introduce new \acp{po} for these cells. This significantly reduces computational complexity. The dynamics of the measurement noise variance is modeled by a Gamma \ac{pdf} $f(\eta_{k}^{(i)} | \eta_{k - 1}^{(i)}) = \mathcal{G}(\eta_{k}^{(i)}; \eta_{k - 1}^{(i)} / c_{\eta}, c_{\eta})$ with $c_{\eta} = 10^2$. At the initial time step, $k = 0$, the prior of noise variances $\eta_{0}^{(i)}, i \in \{1, \dots, 7\}$ is uniform over $[0, 2 \times 10^{-4}]$ and there is no prior information for \ac{po} states, i.e., $N_0 = 0$. The survival probability is set to $p_{\text{s}} = 0.95$, the declaration threshold to $T_{\text{dec}} = 0.5$, and the pruning threshold to $T_{\text{pru}} = 10^{-2}$. The functions $\V{a}^{(i)}(\V{x}) \in \mathbb{C}^M$, $i \in \{1,\dots I\}$ representing the frequency-dependent acoustic pressure field at the array for a source with unit amplitude and state $\V{x}$, can be evaluated using a simulator of the underwater propagation environment \cite{JenKupPor:B11}.

As reference methods, we consider the conventional multi-object tracking method in \cite{MeyKroWilLauHlaBraWin:J18} paired with two different preprocessing stages, namely \ac{mp} \cite{MalZha:J93} and \ac{sbl} \cite{GemNanGerHod:J17, NanGemGerHodMec:J19}. The pairing of \ac{sbl} with conventional multi-object tracking in \cite{MeyKroWilLauHlaBraWin:J18}, was first introduced in \cite{LiLeiVen:J22}. As a result, \ac{mp}  and \ac{sbl}  provide range-depth measurements as the inputs for conventional multi-object tracking. Direct tracking and conventional multi-object tracking are both implemented using particle-based \ac{bp} \cite{IhlMca:09} and their performance is evaluated using the \ac{gospa} \cite{GarWymLarHaiCou:17} metric with cutoff parameter $c = 200$ and order $p = 1$. \Ac{gospa} results are shown in Fig.~\ref{fig:gospa}. In addition, a visualization of the tracking results at time step $k = 200$ is shown in Fig.~\ref{fig:qualitative}. It can be seen that the proposed method outperforms the two reference methods. The performance difference is most significant at $k = 200$. One key reason is that the performance of the conventional two-step methods highly depends on the quality of the detector, which often produces incorrect detections due to side lobes and model mismatch. The proposed direct tracking method can improve performance by avoiding a potentially error-prone detector. Note that the \ac{gospa} error peaks around  $k=90$ and $k=140$ are related to the time steps where the sources do not transmit the aforementioned tones but a different wavefrom. Direct tracking can partly still succeed in tracking both sources during the aforementioned time steps due to its ability to adapt transmit power\vspace{-0mm}.

\acresetall
\section{Conclusion}

This paper proposes a method that makes it possible to directly track multiple objects with minimal data preprocessing. This contrasts conventional tracking methods that use so-called ``point'' measurements as inputs. These point measurements are generated through a preprocessing stage that typically consists of coherent ``matched'' filters to form data cells, e.g., in range and bearing, followed by a detector for the extraction of point measurements.  

For direct tracking, we introduced a new measurement model to describe the data-generating process of the sensor data. Here, the signal amplitude is modeled as random and zero-mean Gaussian distributed with a variance governed by the object's existence variable and transmit power. The resulting hierarchical Bernoulli-Gaussian model has sparsity-promoting behavior that facilitates the separation of signal contributions of closely spaced objects. While the new model is very general and can be applied to a wide range of active and passive tracking problems, for the sake of simplicity, we here focus on the passive tracking of sources that emit tonal signals. For passive tracking, the introduced measurement model supports multiple tonal signals and multiple snapshots of data. Our joint statistical model, consisting of the measurement and state transition model, is represented by a factor graph, laying the foundation for developing a computationally efficient particle-based \ac{bp} method for direct tracking. Performance has been evaluated in a passive acoustic scenario. Future work includes comparisons with methods that integrate ray tracing and probabilistic data association for passive tracking \cite{MeyWil:J21,WatStiTes:C24} and efficient BP-message computations using deterministic or stochastic particle flow \cite{ZhaMey:J24,ZhaKhoMey:C24}. \vspace{0mm}







\ifCLASSOPTIONcaptionsoff
  \newpage
\fi


\renewcommand{\baselinestretch}{1.018}
\bibliographystyle{ieeetr}
\bibliography{IEEEabrv,StringDefinitions,Books,Papers,ref,refBooks}

\begin{thebibliography}{10}

\bibitem{BarWilTia:B11}
Y.~Bar-Shalom, P.~K. Willett, and X.~Tian, {\em {Tracking and Data Fusion: A
  Handbook of Algorithms}}.
\newblock Storrs, CT: Yaakov Bar-Shalom, 2011.

\bibitem{Mah:B07}
R.~Mahler, {\em {Statistical Multisource-Multitarget Information Fusion}}.
\newblock Norwood, MA: Artech House, 2007.

\bibitem{ChaMor:B11}
S.~Challa, M.~R. Morelande, D.~Mu{\v s}icki, and R.~J. Evans, {\em
  {Fundamentals of Object Tracking}}.
\newblock Cambridge, UK: Cambridge University Press, 2011.

\bibitem{MeyKroWilLauHlaBraWin:J18}
F.~Meyer, T.~Kropfreiter, J.~L. Williams, R.~Lau, F.~Hlawatsch, P.~Braca, and
  M.~Z. Win, ``Message passing algorithms for scalable multitarget tracking,''
  {\em Proc. {IEEE}}, vol.~106, pp.~221--259, Feb. 2018.

\bibitem{LiaMey:J23}
M.~Liang and F.~Meyer, ``Neural enhanced belief propagation for multiobject
  tracking,'' {\em {IEEE} Trans. Signal Process.}, vol.~72, pp.~15--30, Sept.
  2023.

\bibitem{Nit:J72}
R.~Nitzberg, ``Constant-false-alarm-rate signal processors for several types of
  interference,'' {\em {IEEE} Trans. Aerosp. Electron. Syst.}, pp.~27--34, Feb.
  1972.

\bibitem{MalZha:J93}
S.~Mallat and Z.~Zhang, ``Matching pursuits with time-frequency dictionaries,''
  {\em {IEEE} Trans. Signal Process.}, vol.~41, pp.~3397--3415, Dec. 1993.

\bibitem{NanGemGerHodMec:J19}
S.~Nannuru, K.~L. Gemba, P.~Gerstoft, W.~S. Hodgkiss, and C.~F.
  Mecklenbräuker, ``Sparse {Bayesian} learning with multiple dictionaries,''
  {\em Signal Processing}, vol.~159, pp.~159--170, June 2019.

\bibitem{YinZhoKra:21}
T.~Yin, X.~Zhou, and P.~Krahenbuhl, ``Center-based {3D} object detection and
  tracking,'' in {\em Proc. CVPR-21}, pp.~11784--11793, June 2021.

\bibitem{TonBar:J98}
S.~Tonissen and Y.~Bar-Shalom, ``Maximum likelihood track-before-detect with
  fluctuating target amplitude,'' {\em {IEEE} Trans. Aerosp. Electron. Syst.},
  vol.~34, pp.~796--809, July 1998.

\bibitem{KimRisGuaRos:21}
D.~Y. Kim, B.~Ristic, R.~Guan, and L.~Rosenberg, ``A {Bernoulli}
  track-before-detect filter for interacting targets in maritime radar,'' {\em
  {IEEE} Trans. Aerosp. Electron. Syst.}, vol.~57, pp.~1981--1991, June 2021.

\bibitem{VoVoPhaSut:10}
B.-N. Vo, B.-T. Vo, N.-T. Pham, and D.~Suter, ``Joint detection and estimation
  of multiple objects from image observations,'' {\em {IEEE} Trans. Signal
  Process.}, vol.~58, pp.~5129--5141, Oct. 2010.

\bibitem{LepRabGla:J16}
A.~Lepoutre, O.~Rabaste, and F.~Le~Gland, ``Multitarget likelihood computation
  for track-before-detect applications with amplitude fluctuations of type
  {S}werling 0, 1, and 3,'' {\em {IEEE} Trans. Aerosp. Electron. Syst.},
  vol.~52, pp.~1089--1107, June 2016.

\bibitem{KroWilMey:21}
T.~Kropfreiter, J.~L. Williams, and F.~Meyer, ``A scalable track-before-detect
  method with {P}oisson/multi-{B}ernoulli model,'' in {\em Proc. FUSION-21},
  pp.~1--8, Nov. 2021.

\bibitem{LiaKroMey:J23}
M.~Liang, T.~Kropfreiter, and F.~Meyer, ``A {BP} method for
  track-before-detect,'' {\em {IEEE} Signal Process. Lett.}, vol.~30,
  pp.~1137--1141, July 2023.

\bibitem{KroWilMey:C24}
T.~Kropfreiter, J.~L. Williams, and F.~Meyer, ``Multiobject tracking for
  thresholded cell measurements,'' in {\em Proc. FUSION-24}, 2024.

\bibitem{DavGar:J24}
E.~S. Davies and {\'A}.~F. Garc{\'\i}a-Fern{\'a}ndez, ``Information exchange
  track-before-detect multi-{Bernoulli} filter for superpositional sensors,''
  {\em {IEEE} Trans. Signal Process.}, vol.~72, pp.~607--621, Jan. 2024.

\bibitem{HanFleRao:J18}
T.~L. Hansen, B.~H. Fleury, and B.~D. Rao, ``Superfast line spectral
  estimation,'' {\em {IEEE} Trans. Signal Process.}, vol.~66, pp.~2511--2526,
  Feb. 2018.

\bibitem{Wil:J15}
J.~L. Williams, ``{Marginal multi-Bernoulli filters: RFS derivation of MHT,
  JIPDA and association-based MeMBer},'' {\em {IEEE} Trans. Aerosp. Electron.
  Syst.}, vol.~51, pp.~1664--1687, Jul. 2015.

\bibitem{KscFreLoe:01}
F.~R. Kschischang, B.~J. Frey, and H.-A. Loeliger, ``Factor graphs and the
  sum-product algorithm,'' {\em {IEEE} Trans. Inf. Theory}, vol.~47,
  pp.~498--519, Feb. 2001.

\bibitem{Loe:04}
H.-A. Loeliger, ``An introduction to factor graphs,'' {\em {IEEE} Signal
  Process. Mag.}, vol.~21, pp.~28--41, Jan. 2004.

\bibitem{ShaKirLi:B02}
Y.~Bar-Shalom, T.~Kirubarajan, and X.-R. Li, {\em {Estimation with Applications
  to Tracking and Navigation}}.
\newblock New York, NY: Wiley, 2002.

\bibitem{IhlMca:09}
A.~Ihler and D.~McAllester, ``Particle belief propagation,'' in {\em Proc.
  AISTATS-09}, vol.~5, pp.~256--263, Apr. 2009.

\bibitem{LiaLeiMey:J24arixv}
M.~Liang, E.~Leitinger, and F.~Meyer, ``Direct multipath-based {SLAM},'' {\em
  arXiv preprint arXiv:2409.20552}, 2024.

\bibitem{BooAbaSchHod:J00}
N.~Booth, A.~Abawi, P.~Schey, and W.~Hodgkiss, ``Detectability of low-level
  broad-band signals using adaptive matched-field processing with vertical
  aperture arrays,'' {\em {IEEE} J. Ocean. Eng.}, vol.~25, pp.~296--313, July
  2000.

\bibitem{GemNanGerHod:J17}
K.~L. Gemba, S.~Nannuru, P.~Gerstoft, and W.~S. Hodgkiss, ``Multi-frequency
  sparse {B}ayesian learning for robust matched field processing,'' {\em J.
  Acoust. Soc. Am.}, vol.~141, pp.~3411--3420, May 2017.

\bibitem{JenKupPor:B11}
F.~B. Jensen, W.~A. Kuperman, M.~B. Porter, and H.~Schmidt, {\em Computational
  Ocean Acoustics}.
\newblock New York, NY: Springer, 2nd~ed., 2011.

\bibitem{LiLeiVen:J22}
X.~Li, E.~Leitinger, A.~Venus, and F.~Tufvesson, ``Sequential detection and
  estimation of multipath channel parameters using belief propagation,'' {\em
  {IEEE} Trans. Wireless Commun.}, vol.~21, no.~10, pp.~8385--8402, 2022.

\bibitem{GarWymLarHaiCou:17}
N.~Garcia, H.~Wymeersch, E.~G. Larsson, A.~M. Haimovich, and M.~Coulon,
  ``Direct localization for massive {MIMO},'' {\em {IEEE} Trans. Signal
  Process.}, vol.~65, pp.~2475--2487, May 2017.

\bibitem{MeyWil:J21}
F.~Meyer and J.~L. Williams, ``Scalable detection and tracking of geometric
  extended objects,'' {\em {IEEE} Trans. Signal Process.}, vol.~69,
  pp.~6283--6298, Oct. 2021.

\bibitem{WatStiTes:C24}
L.~Watkins, P.~Stinco, A.~Tesei, and F.~Meyer, ``A probabilistic focalization
  approach for single receiver underwater localization,'' in {\em Proc.
  FUSION-24}, 2024.

\bibitem{ZhaMey:J24}
W.~Zhang and F.~Meyer, ``Multisensor multiobject tracking with improved
  sampling efficiency,'' {\em {IEEE} Trans. Signal Process.}, vol.~72,
  pp.~2036--2053, 2024.

\bibitem{ZhaKhoMey:C24}
W.~Zhang, M.~J. Khojasteh, and F.~Meyer, ``{Particle flows for source
  localization in 3-D using TDOA measurements},'' in {\em Proc. FUSION-24},
  2024.

\end{thebibliography}

\end{document}